\documentclass[12pt]{article}

\newcommand{\be}{\begin{equation}}
\newcommand{\ee}{\end{equation}}
\newcommand{\bea}{\begin{eqnarray}}
\newcommand{\eea}{\end{eqnarray}}
\newcommand{\nn}{\nonumber}

\input epsf.sty

\begin{document}

\title{Nonlinear translational symmetric
equilibria  relevant to the L-H transition\\}
\author{Ap Kuiroukidis\footnote{Technological Education Institute of Serres, 62124 Serres, Greece}
and G. N. Throumoulopoulos\footnote{University of Ioannina, Association Euratom-Hellenic Republic, Department of  Physics,
GR 451 10 Ioannina, Greece} \footnote{E-mails: kouirouki@astro.auth.gr,$\; \; $gthroum@uoi.gr}}

\maketitle


\begin{abstract}
Nonlinear   $z$-independent solutions to a generalized
Grad-Shafranov equation (GSE)   with up to quartic flux terms in the
free functions and  incompressible plasma flow non parallel to the
magnetic field are constructed quasi-analytically. Through an ansatz
the GSE is transformed to a set of three ordinary differential
equations and a constraint for three functions of the coordinate
$x$, in cartesian coordinates $(x,y)$, which then are solved
numerically. Equilibrium configurations for certain values of the
integration constants are displayed. Examination of their
characteristics  in connection with the impact of nonlinearity
and sheared flow  indicates that these equilibria are consistent with
the L-H transition phenomenology. For flows parallel to the magnetic
field  one equilibrium corresponding to the H-state is potentially
stable in the sense that a sufficient condition for linear stability
is satisfied in an appreciable part of the plasma while another
solution  corresponding to the L-state does not satisfy the
condition. The results indicate that  the sheared flow in
conjunction with the equilibrium nonlinearity play a stabilizing
role.

\end{abstract}


\section{Introduction}

For axisymmetric toroidal plasma equilibria the force-balance
equation and Maxwell's equations reduce to the Grad-Shafranov
equation (GSE) for the poloidal magnetic flux function $\psi$ \cite{shaf}, \cite{grad}.
Analytical solutions to the
GSE are obtained by specifying the plasma pressure and poloidal
current functions of $\psi$, usually in
such a way as to linearize the resulting partial differential
equation,  e.g. \cite{solov}-\cite{shri}. Analytical solutions to
the GSE are very useful for theoretical studies of plasma
equilibrium, transport and stability as well as benchmarks for
numerical codes \cite{mukh}. Also it has been established  in a
variety of magnetic configurations that sheared flows can reduce
turbulence and produce transport barriers, which under certain
conditions can extend to the whole plasma, e.g. \cite{terry}. In
view of a fusion reactor the spontaneous formation of transport
barriers, i.e., those driven by internal processes even in the
absence of external  sources, is of particular interest. For this
reason among others stationary equilibria with plasma flow have been
extensively studied on the basis of generalized GSEs, e.g. \cite{mape1}-\cite{throu4}.
In particular, although complex numerical codes are
extensively used to attempt simulations of the L-H transition, certain
equilibrium considerations in  connection
with this transition  are  helpful, e.g. \cite{sol}-\cite{tsna}.

The simplest known and widely used in various studies, analytical
solution to the GSE, is the Solov\'ev equilibrium \cite{solov}.
Extension of the original Solov\'ev solution, to include the
possibility of sheared flows appeared in \cite{simi}. In other
extensions additional free parameters were introduced to construct
configurations with fusion relevant  plasma boundaries and desirable
values of confinement figures of merit as the safety factor on magnetic axis \cite{freid,shri,ThTa3}.
Most of the solutions are associated with pressure and current
profiles, including up to quadratic terms in the flux function
$\psi $ to linearize the resulting equation
\cite{solov}-\cite{shri}.   Linear equilibria with flow were
constructed in \cite{mape1}-\cite{kuith2} and in Refs. cited
therein. Also, the non linear translational symmetric equilibria of ``cat eyes" and counter
rotating vortices with flow parallel to the magnetic field  were
studied  in \cite{ttp,throu4}. These nonlinear equilibria, however,
are periodic in  one direction ($x$) and therefore the plasma is not
bounded along this direction.

In most of the above cases the axisymmetric equilibria are obtained
as  separable solutions of GSE.
  A novel  non-separable class of
solutions was found in Ref. \cite{kui} describing up-down symmetric
configurations with incompressible flows parallel to the magnetic
field and it was extended recently to include asymmetric
configurations \cite{kuith1} and  flows of arbitrary direction
\cite{kuith2}.  For non parallel flows the  question of the
stability  is usually not considered and this is partly due to the
difficulty of the subject and the absence of a concise criterion.
Few sufficient conditions for linear stability are available only
for parallel  flows \cite{FrVi}-\cite{throu3}. In previous studies
we found that the stability condition of Ref. \cite{throu3} is not
satisfied for the linear equilibria of \cite{apost} and
{\cite{ThTa3}  while it is satisfied within  an appreciable part of
the plasma  for the nonlinear equilibria of \cite{ttp} and
\cite{throu4}.   This led us to the conjecture that the equilibrium
nonlinearity may act synergetically with the  sheared flow to
stabilize the plasma.


Aim of the present study is to construct certain  two dimensional
nonlinear up-down symmetric equilibria with incompressible flow of
arbitrary direction in  $z$-independent geometry.   They are more
pertinent to a magnetically confined plasma than those of Refs.
\cite{ttp} and \cite{throu4} because the plasma is bounded on the
poloidal plane. Another reason for considering  translational
symmetry is the many free physical and geometrical parameters
involved in connection with the flow amplitude,  direction and
shear, equilibrium nonlinearity, symmetry and toroidicity. Thus, in
the presence of nonlinearity one first could exclude toroidicity.
The study is performed quasi-analytically through a non separable
ansatz under which the GSE is transformed to a set of three ordinary
differential equations and a constraint for three functions.  The
solutions give nested magnetic surfaces and their characteristics
are studied by means of certain equilibrium quantities and
confinement figures of merit as the safety factor, electric field
and ${\bf E} \times {\bf B}$ velocity shear.   Also, for parallel
flows the linear stability is considered by means of the
aforementioned sufficient condition \cite{throu3}. The results are
in qualitative agreement with phenomenological characteristics of an
edge transport barrier, confirm relevant scenarios \cite{terry},
\cite{simi}  and support the above conjecture.

The organization of the paper is as follows: In the first section we
briefly review the general setting for the equations of
incompressible flow with translational symmetry together with the
generalized GSE. In Section II the proposed ansatz and the resulting
equations are presented which then are integrated numerically. In
section III we consider the solutions for certain values of the
various parameters and integration constants and discuss the most
important equilibrium properties. In section IV the criterion for
linear stability is applied to the equilibria constructed for
parallel flows. Section V summarizes the study
and briefly proposes  potential extensions. \\
\\


\section{Translational Symmetric Equilibria with flow}\

The equilibrium of a cylindrical plasma with
incompressible flow and arbitrary cross-sectional shape, satisfies
 \cite{throu1}, \cite{simi}, \bea \label{eqpsi}
(1-M_{p}^{2})\nabla^{2}\psi-\frac{1}{2}(M_{p}^{2})^{'}|\nabla\psi|^{2}
+\frac{d}{d\psi}\left(\mu_0 P_{s}+\frac{B_{z}^{2}}{2}\right)=0 \eea
for the poloidal magnetic flux function $\psi$. Here, $M_p(\psi)$,
$P_s(\psi)$, $\rho(\psi)$ and  $B_{z}(\psi)$ are respectively the
poloidal Alfv\'en Mach function,  pressure  in the absence of flow,
density and magnetic field parallel to the symmetry axis $z$, which
are surface quantities. Because of the symmetry, the equilibrium
quantities are $z-$independent and the axial velocity $v_{z}$ does
not appear explicitly in (\ref{eqpsi}). Derivation of (\ref{eqpsi})
is based on the following two steps: First, express the divergence
free fields in terms of scalar quantities as  \begin{eqnarray*}
{\bf B}&=&B_{z}\nabla z+\nabla z\times \nabla \psi\\
{\mu_0 \bf j}&=&\nabla^{2}\psi\nabla z-\nabla z\times \nabla B_{z}\\
\rho{\bf v}&=&\rho v_{z}\nabla z+\nabla z\times \nabla F
\end{eqnarray*} and the electric field by ${\bf E}=-\nabla \Phi $.
Second, project the momentum equation, $\rho\left( {\bf v\cdot
\nabla}\right) {\bf v}= {\bf j} \times {\bf B} -\nabla P$, and Ohm's
law, ${\bf E}+{\bf v}\times {\bf B}=0$,  along the symmetry
direction $z$, {\bf B} and $\nabla \psi$. The projections yield four
first integrals in the form  of surface quantities (two out of which
are $F(\psi)$ and $\Phi(\psi)),  $ Eq. (\ref{eqpsi}) and the
Bernoulli relation for the pressure \bea \label{piesn}
P=P_{s}(\psi)-\frac{1}{2 \mu_0}M_{p}^{2}(\psi)|\nabla\psi|^{2} \eea
Because of the flow $P$ is not a surface quantity.
Also the density
becomes  surface quantity because of incompressibility and
$M_{p}^{2}(\psi)=(F^{'}(\psi))^{2}/(\mu_0\rho)$.  Five of the
surface quantities, chosen here to be    $P_s,\; \rho, \; B_{z},\;
M_p^2$ and $v_{z}$, remain arbitrary.

 Using the transformation
\bea \label{transf} u(\psi)=\int_{0}^{\psi}[1-M_{p}^{2}(g)]^{1/2}dg,\; \;
(M_{p}^{2}<1) \eea Eq. (\ref{eqpsi}) is transformed to \bea
\label{equ} \nabla^{2}u+\frac{d}{du}\left(\mu_0
P_{s}+\frac{B_{z}^{2}}{2}\right)=0 \eea Note that transformation (\ref{transf})
does not affact the magnetic surfaces, it just relabels them.  Eq.  (\ref{equ}) is
identical in form with the static equilibrium equation. In the
present study we assign the free function term in (\ref{equ}) as
 \bea
\label{ansatz1} \left(\mu_0 P_{s}+\frac{B_{z}^{2}}{2}\right)=
c_{0}+c_{1}u+c_{2}\frac{u^{2}}{2}+c_{3}\frac{u^{3}}{3}+
c_{4}\frac{u^{4}}{4} \eea where $c_0, c_1, \ldots, c_4$ are free
parameters.

\section{Proposed Ansatz}\

We use  Eq. (\ref{ansatz1}) into Eq. (\ref{equ}),  employ the ansatz
\bea \label{flux}
u=\frac{N_{1}(x)y^{2}+f(x)D_{0}(x)}{y^{2}+D_{0}(x)} \eea and equate
 the nominator of the resulting equation to zero. From
the $y^{6}-$terms we obtain (a prime denotes derivative with respect
to $x$) \bea \label{n1} N_{1}^{''}
+c_{1}+c_{2}N_{1}+c_{3}N_{1}^{2}+c_{4}N_{1}^{3}=0 \eea From the
$y^{0}-$terms we obtain the constraint $C_{s}=0$, where \bea
\label{constr}
C_{s}=2(N_{1}-f)+D_{0}[c_{1}+c_{2}f+c_{3}f^{2}+c_{4}f^{3}]=0 \eea
The $y^{4}$ and $y^{2}-$terms, after rearrangement yield  \bea
\label{f} f^{''}+2(N_{1}-f)\left(\frac{D_{0}^{'}}{D_{0}}\right)^{2}-
\frac{8(N_{1}-f)}{D_{0}}+c_{4}(N_{1}-f)^{3}=0\nn \\
\eea
and
\bea
\label{d0}
D_{0}^{''}+2\frac{(N_{1}^{'}-f^{'})}{(N_{1}-f)}D_{0}^{'}+
2\frac{(D_{0}^{'})^{2}}{D_{0}}-6+\nn \\
+c_{3}D_{0}(N_{1}-f)+3c_{4}D_{0}N_{1}(N_{1}-f)=0
\eea
Eq. (\ref{n1}) is solved using the tanh method \cite{mal}, a method of solving
non linear differential equations, which also employed  in \cite{kha}.
We have two solutions. The first is
$N_{1}(x)=a_{0}+a_{1}tanh(vx)$, where
\bea
c_{1}+c_{2}a_{0}+c_{3}a_{0}^{2}+c_{4}a_{0}^{3}&=&0\nn \\
c_{2}+c_{3}(2a_{0})+c_{4}(3a_{0}^{2})&=&2v^{2}\nn \\
c_{3}+c_{4}(3a_{0})&=&0\nn \\
c_{4}(a_{1}^{2})&=&-2v^{2}
\eea
and the second is $N_{1}(x)=a_{0}+a_{1}/cosh(vx)$,
where
\bea
c_{1}+c_{2}a_{0}+c_{3}a_{0}^{2}+c_{4}a_{0}^{3}&=&0\nn \\
c_{2}+c_{3}(2a_{0})+c_{4}(3a_{0}^{2})&=&-v^{2}\nn \\
c_{3}+c_{4}(3a_{0})&=&0\nn \\
c_{4}(a_{1}^{2})&=&2v^{2}
\eea


\section{Solutions and equilibrium properties}\

We have solved numerically Eqs. (\ref{constr}), (\ref{f}) and
(\ref{d0}). Using the first of the solutions for $N_{1}$, namely the
tanh solution, we obtained the equilibrium of Fig. 1. We have used
$a_{0}=1.1$, $a_{1}=2.5$, $v=0.6$ and in Eq. (\ref{ansatz1})
$c_{0}=2.588,$ $c_{1}=-0.638$, $c_{2}=0.302$, $c_{3}=0.38$,
$c_{4}=-0.115$. The boundary flux surface corresponds to
$u_{b}=0.11$ while on the magnetic axis $u_{a}=0.$ The constraint
was kept close to zero for the whole of the integration process and
we got an average value of $|C_{s}|$ equal to 0.10.
Given the nonlinearity and complexity of
the method this implies that the solution is indeed acceptable.
Simple quadratic fitting
gives\\
$f=1.272x^{2}+0.049x+0.001$ and $D_{0}=1.488x+3.3$.

Using the second of the solutions for $N_{1}$, namely the cosh
solution, we obtained the equilibrium of Fig. 2. We have used
$a_{0}=1.0$, $a_{1}=-1.6$, $v=1.15$ and in Eq. (\ref{ansatz1})
$c_{0}=2.588 ,$ $c_{1}=0.289$, $c_{2}=1.777$, $c_{3}=-3.099$,
$c_{4}=1.033$. The boundary flux surface corresponds to
$u_{b}=-0.05$ while on the magnetic axis $u_{a}=0.$ The constraint
was kept close to zero for the whole of the integration process and
the average value of $|C_{s}|$ was 0.01.
Simple quadratic fitting gives $f=-0.542x^{2}+0.009x$ and
$D_{0}=0.994x^{2}+3.3$.

Here instead of the velocity $v_z$ we have used the axial Mach
function, $M_{z}^{2}(u)=v_{z}^{2}/(B_{z}^{2}/(\mu_0 \rho)$, and the
approximation  $M_z^2 \approx M_{p}^{2}=(F^{'})^{2}/(\mu_0 \rho)$ in
relation to the tokamak scaling $B_p \approx 0.1 B_z$ and $v_p \approx
0.1 v_z$. In addition, to completely construct the equilibrium we
have made the following choices \bea
 M_{p}^2&=&C_p(u-u_b)^n(u_a-u)^m  \label{assign1} \\ C_p& =& M_{pa}
 \left\lbrack\frac{m (u_a-u_b)}{m+n}\right\rbrack^{-m}
\left\lbrack \frac{n (u_a-u_b)}{m+n}\right\rbrack^{-n} \nn \\
M_{z}^2&=&C_z(u-u_b)^n(u_a-u)^m  \label{assign2} \\ C_z& =& M_{za}
\left\lbrack\frac{m (u_a-u_b)}{m+n}\right\rbrack^{-m}
\left\lbrack \frac{n (u_a-u_b)}{m+n}\right\rbrack^{-n} \nn \\
B_{z}^2&=&B_{z0}^2\left\lbrack 1-\gamma\left(1-\frac{u}{u_{b}}\right)\right\rbrack  \label{assign3} \\
\rho&=&\rho_{a}\left(1-\frac{u}{u_{b}}\right)^{\lambda} \label{assign4}
\eea
 for the
poloidal Mach function, axial Mach function, axial magnetic field  and
density, respectively, with $B_{z0}=2.24 $ T,
 $\rho_a=4\times 10^{-7}$  Kgr/m$^{3}$ $\gamma=0.02$, $u_a=0$, $u_b=0.11$
W m$^2$, (with the subscripts a and b  indicating the magnetic axis and boundary  respectively),
$\lambda=0.5, $ $m=9 n$ and
$M_{za}=1.1M_{pa}$ with various values of the parameters $M_{pa}$ and $n$.  Here,
Eqs. (\ref{assign1}) and (\ref{assign2}) can decribe Mach functions localized
in the edge plasma region in connection with the L-H transition (in particular flows
localized nearly in the one tenth of the exterior  plasma will be considerded
as it is shown in Fig. 3);    Eq. (\ref{assign3}) represents a diamagnetic
$B_z(u)$   (Fig. 4). Then,  (\ref{piesn}) and (\ref{ansatz1}) imply a pressure
peaked on axis (Fig. 5).

Furthermore we have examined certain  equilibrium characteristics by
means of the safety factor, magnetic shear, axial current density,
radial electric field and  ${\bf E}\times {\bf B} $ velocity shear, , and found the following results.
\begin{enumerate}
\item  The safety factor for both solutions  shown  in Figs.
6,7  is slightly affected by the  flow. Also, the flow affects
slightly the magnetic shear given by $s(u)=2(V/q)(dq/dV)$  as it can
be seen in  Fig. 8 for Equilibrium 1. A similar plot holds for
Equilibrium 2.
\item The
 radial electric
field for the two solutions has an  extremum in the edge region
which increases with flow (Figs. 9 and 10). The position of the
extremum, however, is nearly unaffected by the flow. These
characteristics are indicative that the solutions may be relevant to
the L-H transition as discussed in  \cite{simi} where a similar
behavior  of the electric field was found  (Fig. 3 therein).
\item The
${\bf E}\times {\bf B}$ velocity shear which is believed to play a
role in the transitions to improved confinement regimes of
magnetically confined plasmas is  given by \bea \omega_{E\times B}=
\left|\frac{d}{dr}\left[\frac{{\bf E}\times {\bf
B}}{B^{2}}\right]\right| \eea where $r$ is the length variable
normal to the magnetic surfaces. For Equilibrium 1 it is plotted in
Fig. 11;  a similar plot holds for Equilibrium 2. $\omega_{E\times
B}$ is increased by the flow in the edge region outer from the local
minimum while it remains nearly unaffected in the central region.
This is another
 indication supporting the relevance of the solutions to the L-H transition.
 \item  The flow makes the
 axial (``toroidal") current density  profile hollow as shown
 in Fig. 12  for Equilibrium 1. (A similar $j_{tor}$ profile is found for Equilibrium 2.)
 The larger the flow is the stronger the hollowness.  Hollow $j_{tor}$ profiles are usually related
 to the formation of internal transport barriers in tokamaks. However, despite of this characteristic  and
 the fact that $\omega_{E\times B}$ becomes maximum on the magnetic axis (Fig. 11) it is
 unlikely that the present equilibria
 are related to internal transport barriers because the safety factor is monotonically increasing from
 the magnetic axis to the plasma edge (Figs. 6, 7). According to observations in tokamaks, e.g  \cite{devr} for JET
 and    \cite{shmc} for DIII-D,
 it is  the reversed magnetic shear
 which  plays a role in triggering the ITBs development.
 Also, as can be seen in Figs. 6 and 7  the flow makes the central $q$-values lower.
\end{enumerate}

\section{Stability consideration}\

We now consider the important issue of the stability of the
solutions constructed in Section IV with respect to small linear MHD
perturbations by applying the  sufficient condition of Ref.
\cite{throu3}. This condition states that a general steady state of
a plasma of constant density and incompressible flow parallel to
$\bf B$ is linearly stable to small three-dimensional perturbations
if the flow is sub-Alfv\'enic ($M^2<1$) and $A\geq 0$, where $A$ is
given below by (\ref{cond}). Consequently, using henceforth
dimensionless quantities we set $\rho=1$. Also, for parallel flows
(${\bf v}=M{\bf B}$) it holds $M_p\equiv M_z\equiv M $. In fact if
the density is uniform at equilibrium it remains so at the perturbed
state because of incompressibility \cite{TaTh2}.
 In the $u$-space for axisymmetric
 equilibria
 $A$ assumes the form
 \bea \label{cond} A&=&-{\bar
g}^{2} \left[\frac{}{}({\bf j}\times \nabla u)\cdot ({\bf
B}\cdot\nabla)\nabla u+\right.
\nn \\
&+&\left(\frac{M_{p}^{2}}{2}\right)^{'}\frac{|\nabla u|^{2}}{(1-M_{p}^{2})^{3/2}}
\left\{\frac{}{}\nabla u\cdot \nabla (B^{2}/2)+\right. \nn \\
&+&\left.\left.{\bar g}\frac{|\nabla
u|^{2}}{(1-M_{p}^{2})^{1/2}}\frac{}{}\right\}\right] \eea
with
$$\bar{g}:=\frac{P_{s}^{'}(u)-(M_{p}^{2})^{'}B^{2}/2}{1-M_{p}^{2}}$$ Symbolic
computation of $A$ over a wide rage of parametric values led to the
following results:
\begin{enumerate}
\item  Equilibrium 1 is not  satisfied, since $A<0$ everywhere, while
Equilibrium 2 is satisfied in an appreciable part of the plasma
region. However, it is noted  that since the stability condition is
necessary, $A<0$ does not imply that an equilibrium is unstable. An
example of the sign of $A$ for Equilibrium 2 is given in the
three-dimensional plot of Fig. 13. Also, profiles of $A$ in the
middle-plane $y=0$ for a static and a stationary equilibrium are
shown in Fig. 14.
\item Increase of $M_{pa}$ makes  $A$ more positive in the edge region as
can be seen in the example of Fig. 14. A similar impact on A has the
 flow shear parameter $n$ (Eq. (\ref{assign1}))
 as can be seen in Fig. 15 showing  the profile
 of $\delta A= A(y=0, n=2)-A(y=0, n=1)$.
 \item The equilibrium nonlinearity in connection with the parameters
 $c_3$ and $c_4$ has a  stabilizing effect in the edge region as shown in the
 example of Fig. 16 plotting the profile of the difference $\delta A $ between a nonlinear
 and a linear Equilibrium 2.
 \end{enumerate}
 According to the above results and the believe  that the sheared flow is developed
 during the L-H transition we conjecture that a static Equilibrium 1 could correspond
 to the L state and a stationary Equilibrium 2 with ${\bf E} \times {\bf B} /B^2$ velocity shear
 to the H state. In a quasistatic evolution approximation the
 plasma could then evolve through successive states with increased sheared flow (increasing values
 of the parameters $M_{pa}$ and $M_{za}$ and most importantly increasing values of the shearing
 parameters $m$ and $n$).

\begin{figure}[h!]
\label{sxnma1}
\centerline{\mbox {\epsfxsize=12.0cm \epsfysize=10.0cm
\epsfbox{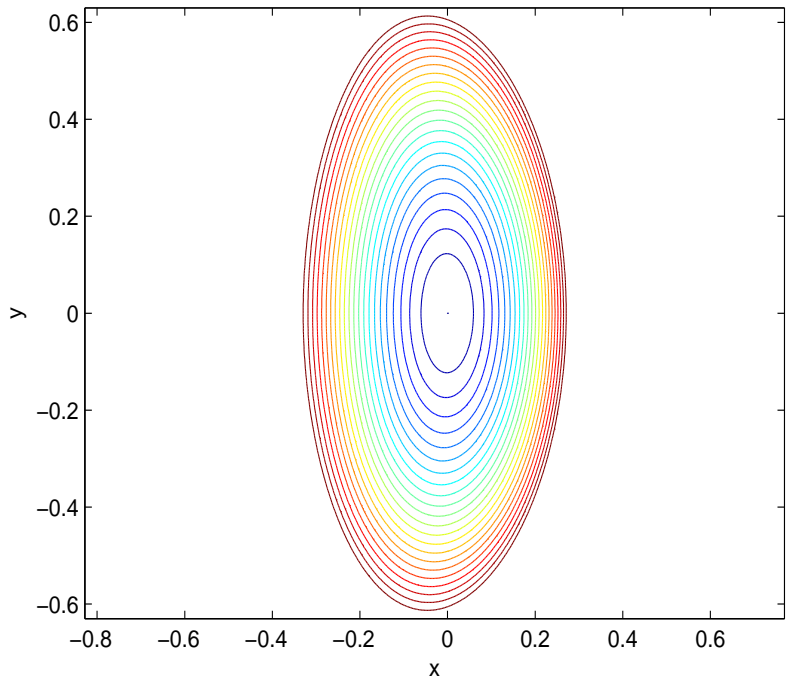}}}
\caption{Equilibrium 1. The bounding flux
surface corresponds to $u_{b}=0.11,$ with
$u_{a}=0$, for the magnetic axis. For this equilibrium
 the average value of $|C_{s}|$ is 0.10.}
\end{figure}

\begin{figure}[h!]
\label{sxnma2}
\centerline{\mbox {\epsfxsize=12.0cm \epsfysize=10.0cm
\epsfbox{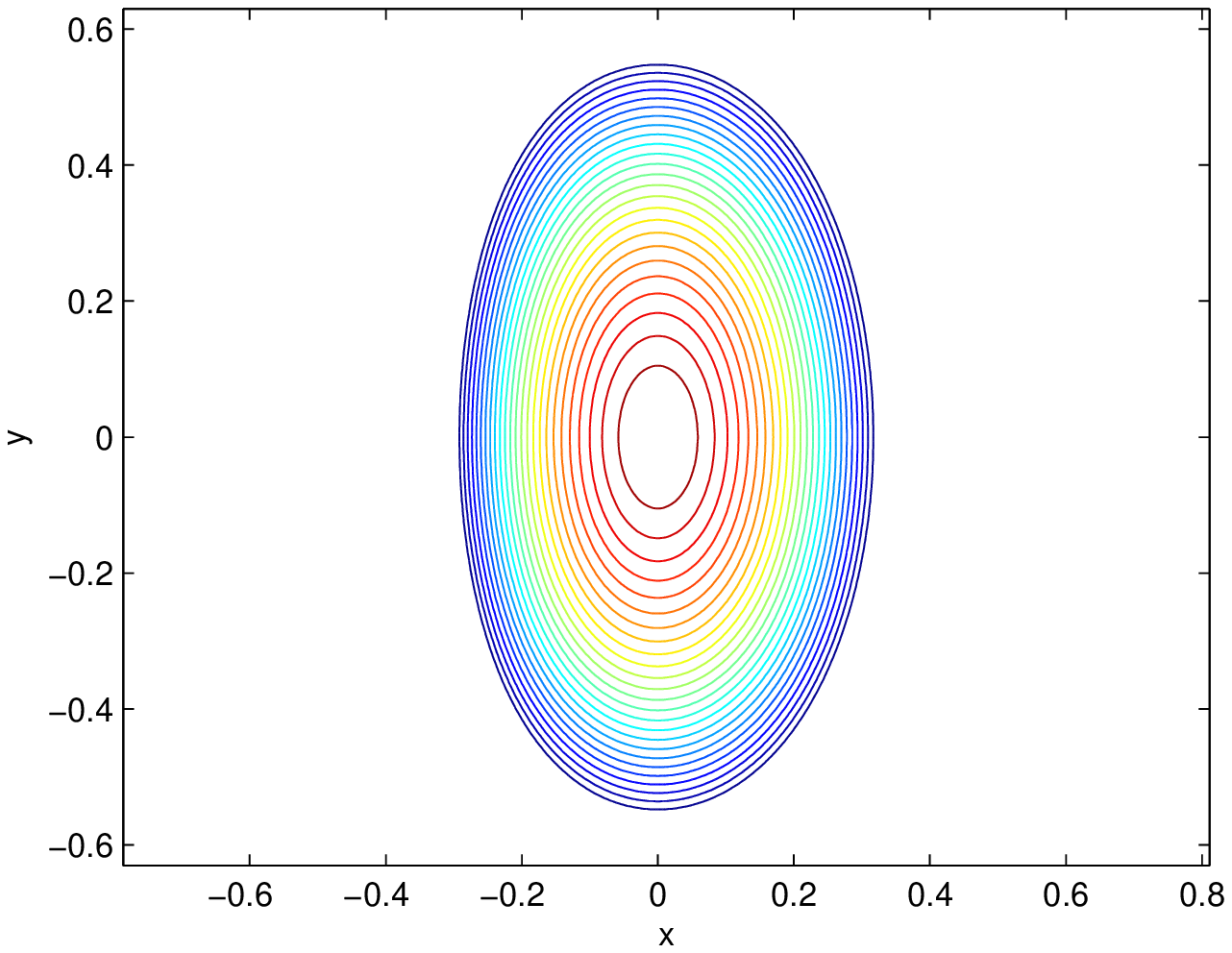}}}
\caption{Equilibrium 2. The bounding flux
surface corresponds to $u_{b}=-0.05,$ with
$u_{a}=0$, for the magnetic axis. For this equilibrium
the average value of $|C_{s}|$ is 0.10.}
\end{figure}

\begin{figure}[h!]
\label{sxnma3a}
\centerline{\mbox {\epsfxsize=12.0cm \epsfysize=10.0cm
\epsfbox{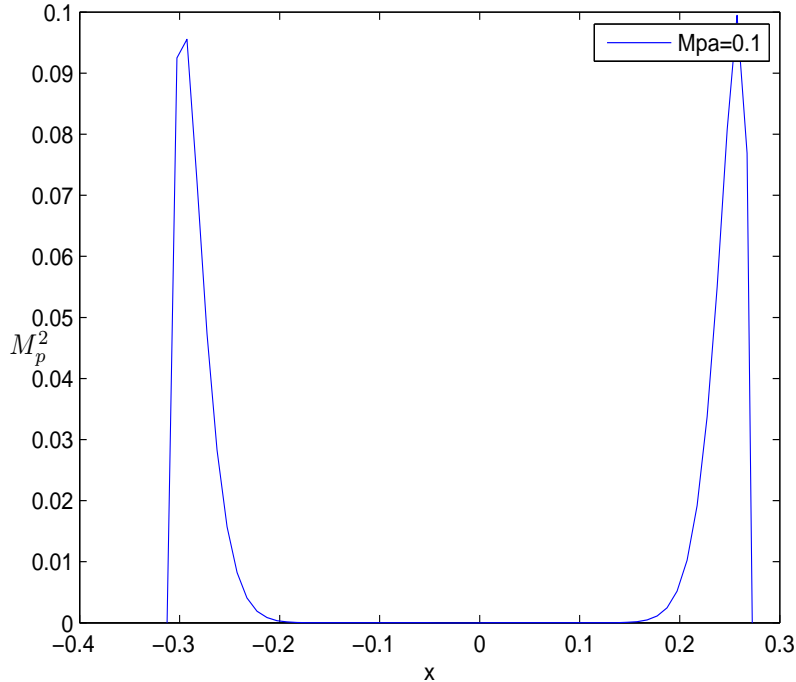}}}
\caption{L-H transition-like  Mach function  in connection with
Eq. (\ref{assign1}) with $n=1$ and a maximum localized at a distance
from the boundary as large as the on tenght  of the minor radius.}
\end{figure}

\begin{figure}[h!]
\label{sxnma3b}
\centerline{\mbox {\epsfxsize=12.0cm \epsfysize=10.0cm
\epsfbox{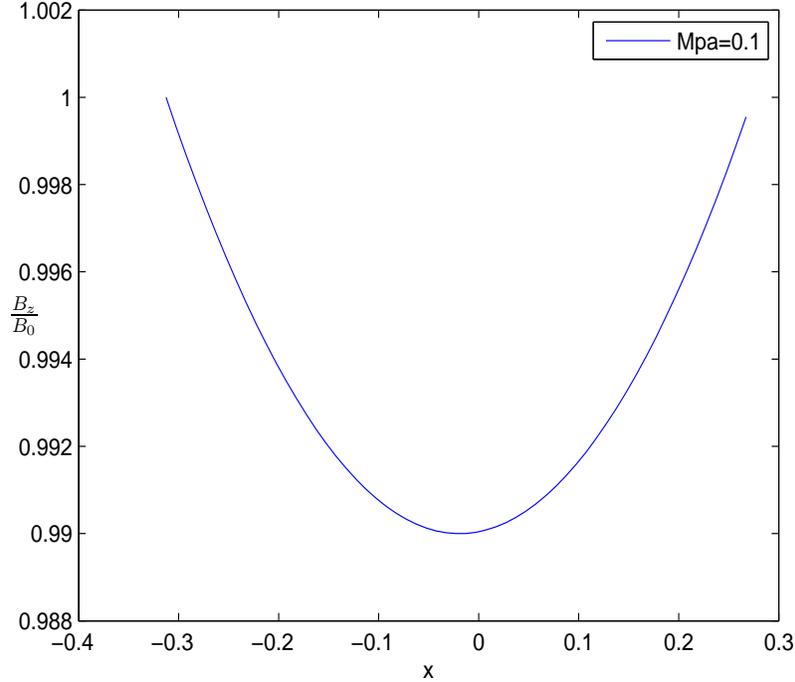}}}
\caption{Typical diamagnetic axial magnetic field profile normalized  with respect to the
value at the magnetic axis  in connection with Eq. (\ref{assign3}) .}
\end{figure}

\begin{figure}[h!]
\label{sxnma3c}
\centerline{\mbox {\epsfxsize=12.0cm \epsfysize=10.0cm
\epsfbox{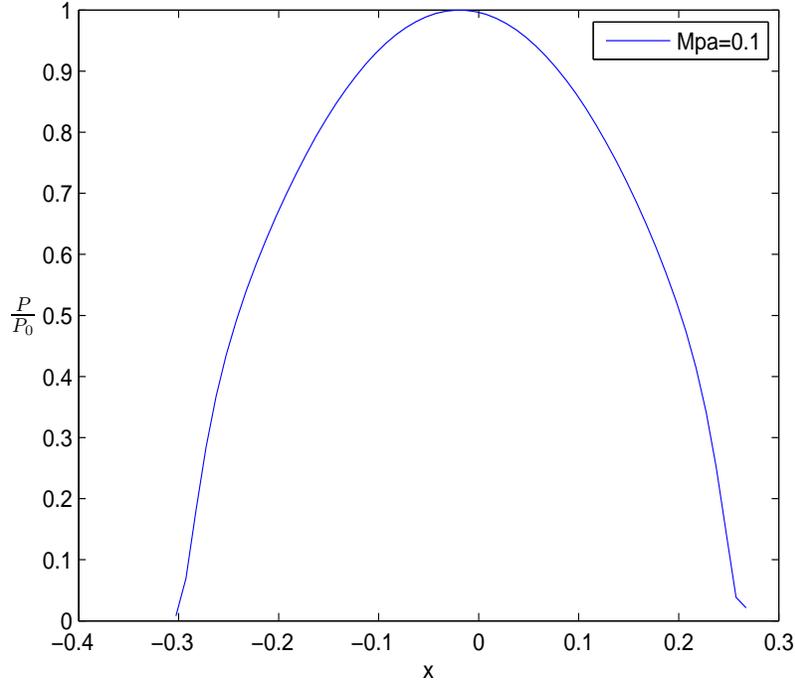}}}
\caption{Pressure profile at $y=0$ for the Equlibrium 1.}
\end{figure}

\begin{figure}[h!]
\label{sxnma4}
\centerline{\mbox {\epsfxsize=12.0cm \epsfysize=10.0cm
\epsfbox{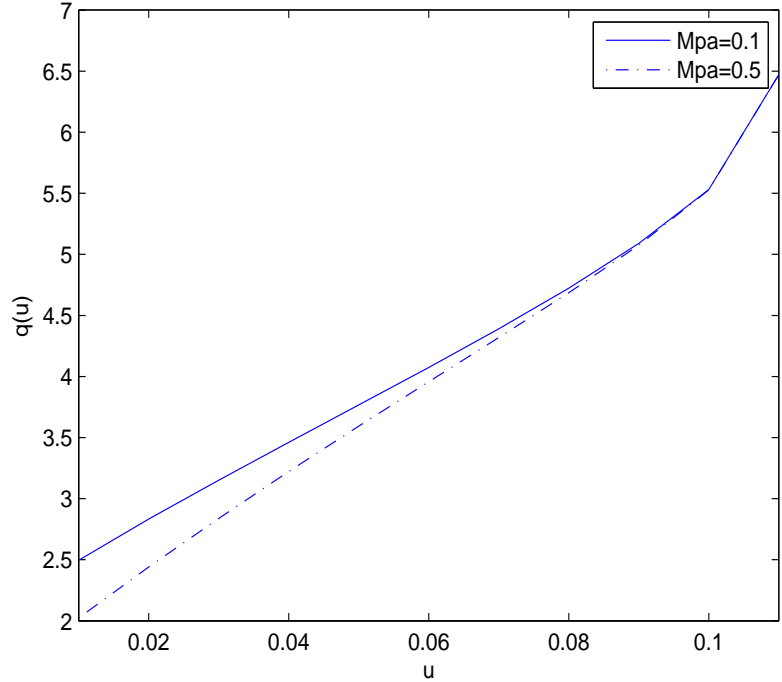}}}
\caption{Safety factor for the solution of Equilibrium 1.}
\end{figure}

\begin{figure}[h!]
\label{sxnma5}
\centerline{\mbox {\epsfxsize=12.0cm \epsfysize=10.0cm
\epsfbox{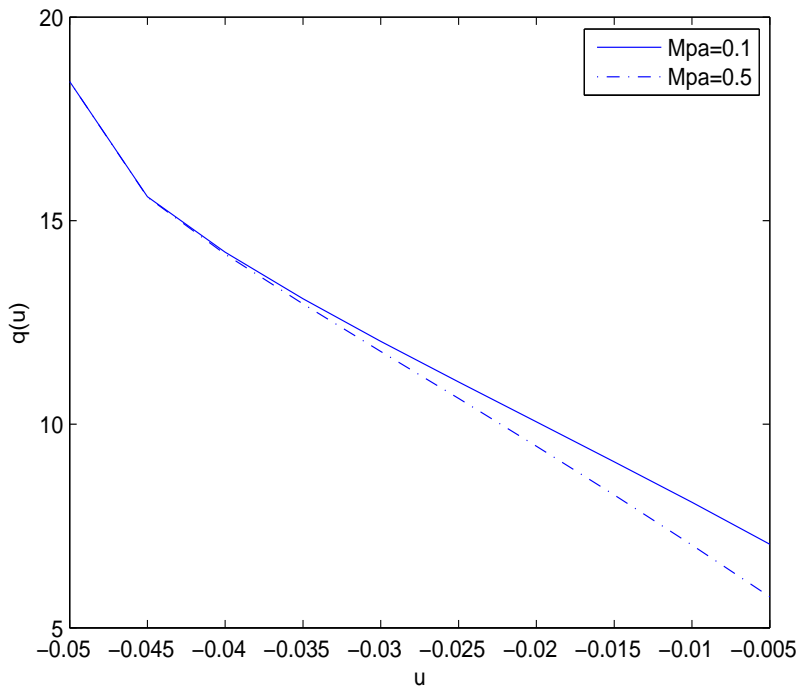}}}
\caption{Safety factor for the solution of Equilibrium 2.
Notice that the outer boundary surface corresponds to
$u=-0.05$ while the magnetic axis to $u=-0.005$}
\end{figure}

\begin{figure}[h!]
\label{sxnma6}
\centerline{\mbox {\epsfxsize=12.0cm \epsfysize=10.0cm
\epsfbox{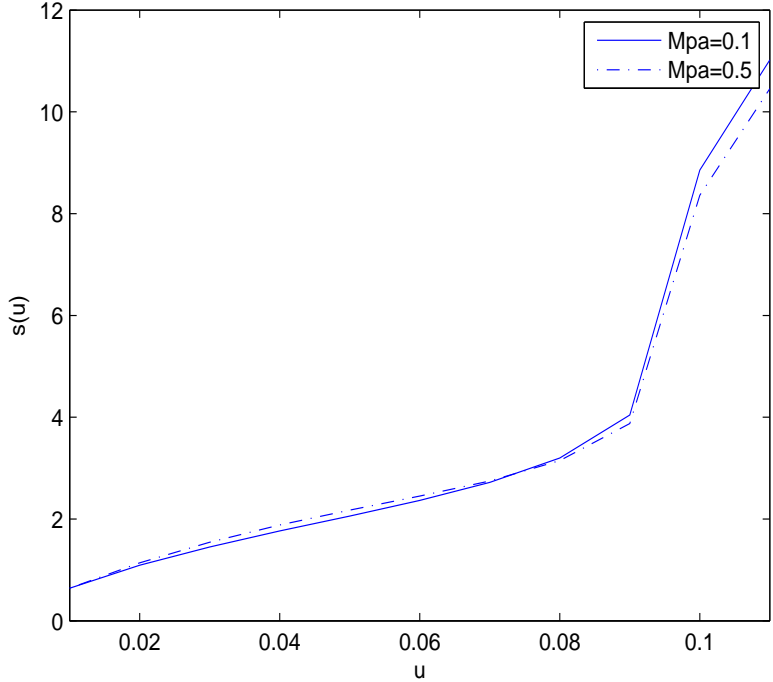}}}
\caption{Magnetic Shear for the solution of Equilibrium 1. It is
slightly affected by the presence of flow. A similar plot holds for
Equilibrium 2.}
\end{figure}

\begin{figure}[h!]
\label{sxnma7}
\centerline{\mbox {\epsfxsize=12.0cm \epsfysize=10.0cm
\epsfbox{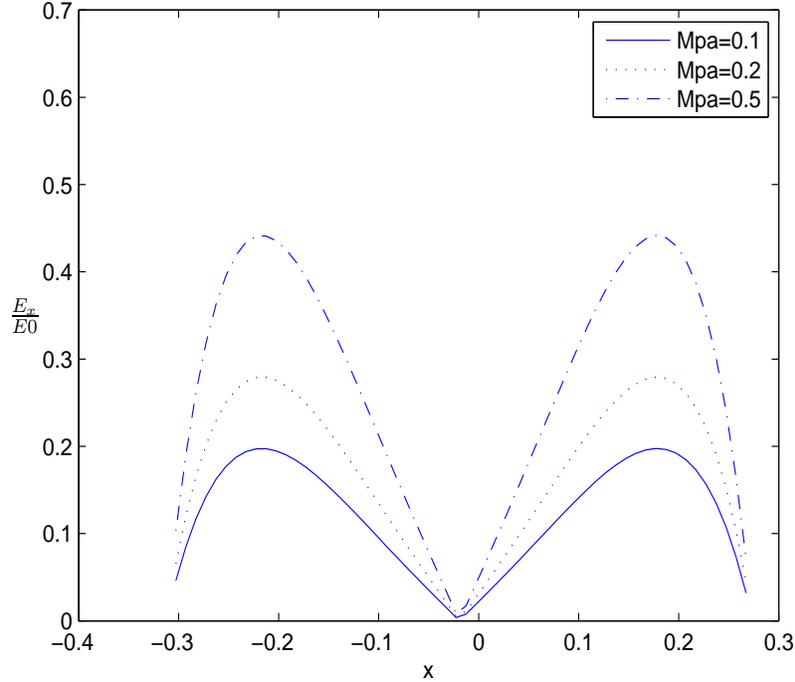}}}
\caption{Normalized electric field with respect to $E_0=280$ kV/m for the solution of Equilibrium 1. The
extremum of the electric field increases with flow, the position
of the extremum however is not significantly affected by it.}
\end{figure}

\begin{figure}[h!]
\label{sxnma8}
\centerline{\mbox {\epsfxsize=12.0cm \epsfysize=10.0cm
\epsfbox{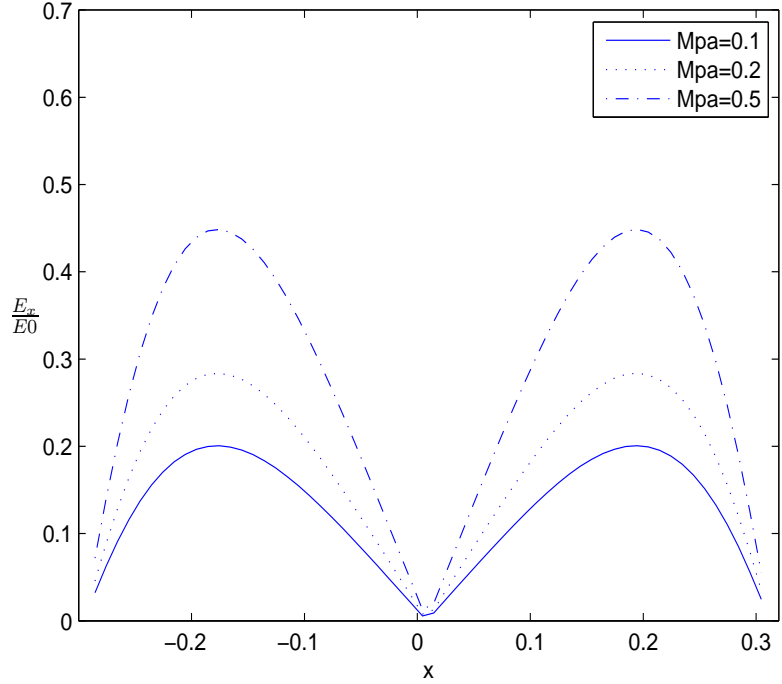}}}
\caption{Electric field for the solution of Equilibrium 2. The
extremum of the electric field increases with flow, the position
of the extremum however is not significantly affected by it.}
\end{figure}

\begin{figure}[h!]
\label{sxnma9}
\centerline{\mbox {\epsfxsize=12.0cm \epsfysize=10.0cm
\epsfbox{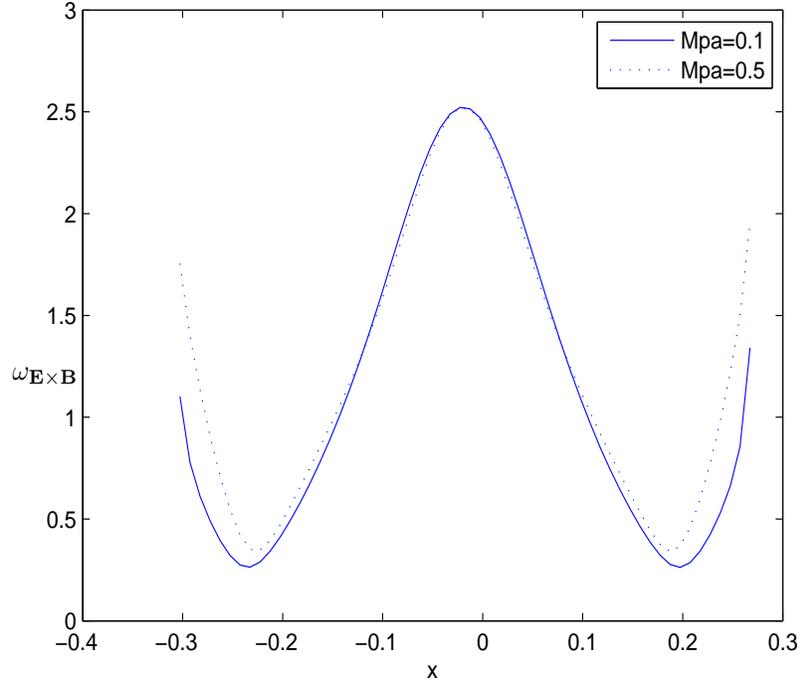}}}
\caption{The ${\bf E}\times {\bf B}$ velocity shear, for the
solution of Equilibrium 1. It increases slightly with the presence
of flow in the edge region. Similar plot holds for Equilibrium 2.}
\end{figure}

\begin{figure}[h!]
\label{sxnma10}
\centerline{\mbox {\epsfxsize=12.0cm \epsfysize=10.0cm
\epsfbox{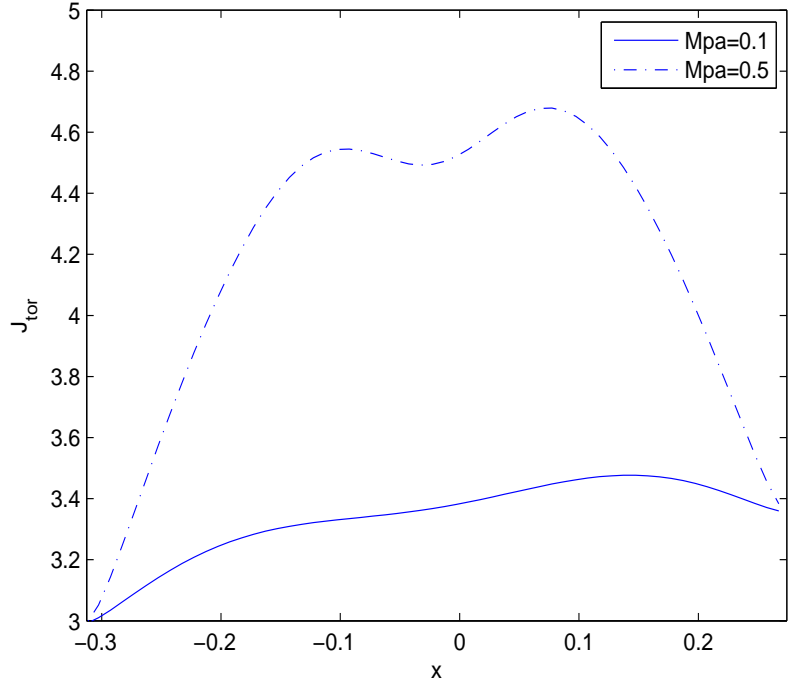}}}
\caption{Axial current density in MA as a function of the
flow parameter $M_{pa}$, for the solution of Equilibrium 1. As the
flow increases a hollow profile in the core of the equilibrium
appears and it becomes larger for larger values of the flow.}
\end{figure}


\begin{figure}[h!]
\label{sxnma12}
\centerline{\mbox {\epsfxsize=12.0cm \epsfysize=10.0cm
\epsfbox{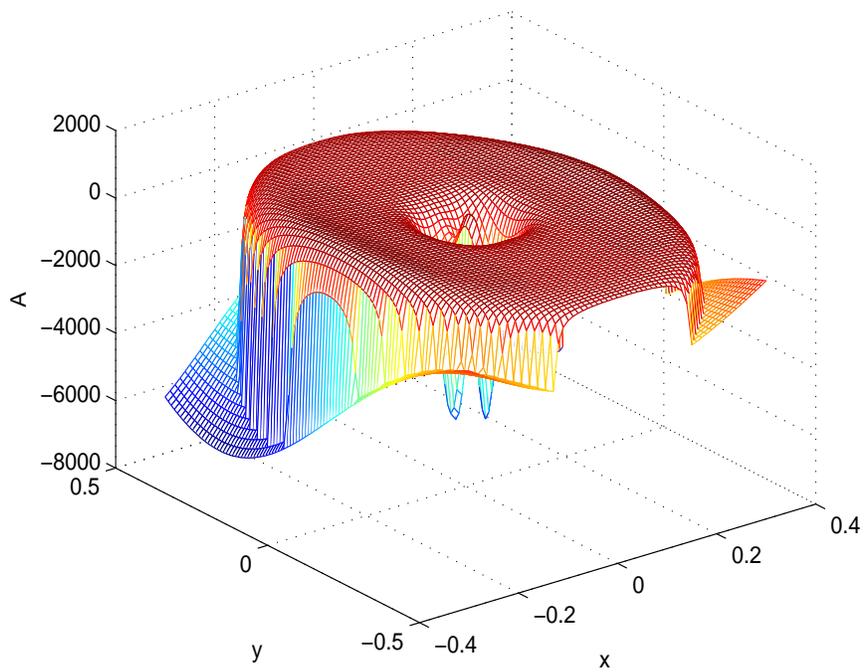}}}
\caption{Stability function for the solution of Equilibrium 2. For
most part of the equilibrium is positive and assumes negative values
only in the core of the equilibrium.}
\end{figure}

\begin{figure}[h!]
\label{sxnma13}
\centerline{\mbox {\epsfxsize=12.0cm \epsfysize=10.0cm
\epsfbox{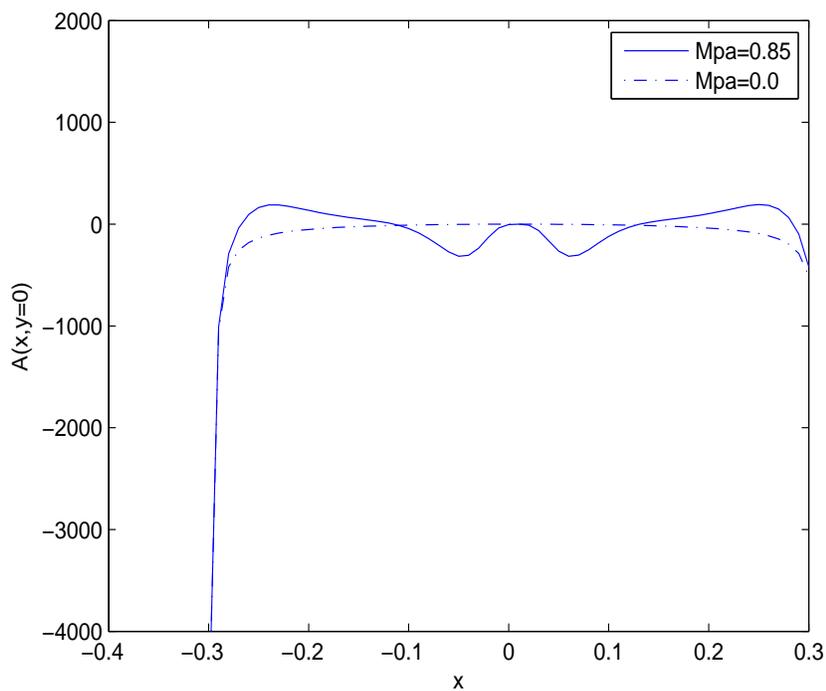}}}
\caption{Plot of the stability function A, at y=0, for the second
equilibrium for nonzero values of the nonlinearity parameters
c3=-3.099, c4=1.033. Increasing the flow parameter $M_{pa} $appears
to improve stability for most part on the  middle-plane except for
the canter of the equilibrium}
\end{figure}

\begin{figure}[h!]
\label{sxnma14}
\centerline{\mbox {\epsfxsize=12.0cm \epsfysize=10.0cm
\epsfbox{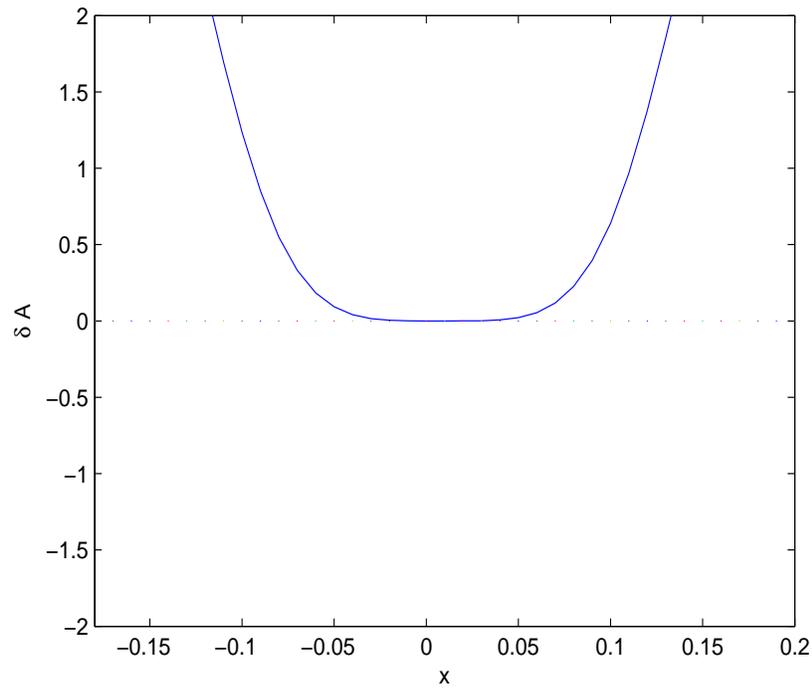}}}
\caption{Plot of the deference $\delta A=A(n=...)-A(n=...)$ for
$c3=-3.099$, $c4=1.033$ and  $Mpa=0.1$ clearly indicating that  the
stability is improved ($\delta A >0$) at the external  part of the
middle-plane $y=0$ as the flow-shear parameter $n$ increases. The
dotted line represents  the $x$-axis. }
\end{figure}

\begin{figure}[h!]
\label{sxnma15}
\centerline{\mbox {\epsfxsize=12.0cm \epsfysize=10.0cm
\epsfbox{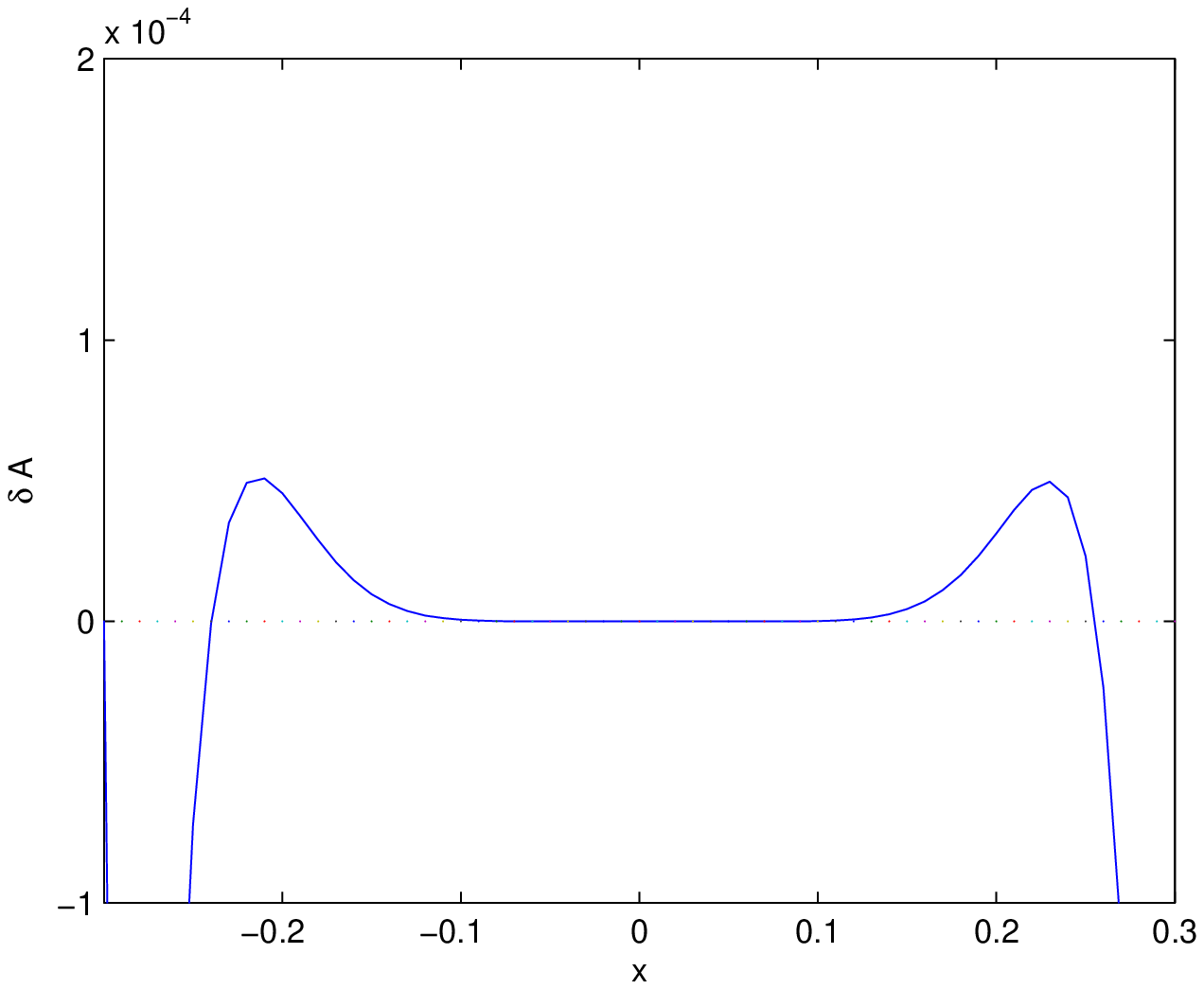}}}
\caption{Plot of the  profile $\delta A=A(c_3=-3.099,
c_4=1.033)-A(c_3= c_4=0)$ at $y=0$ indicating that the nonlinearity
has a  stabilizing effect ($\delta A >0$) in the edge region}
\end{figure}


\section{Summary}

Two classes of solutions of nonlinear  two dimensional magnetohydrodynamic
equilibria for bounded magnetically confined plasmas with sheared
incompressible non parallel flows have been constructed in
cylindrical ($z$-independent) geometry. The equilibria hold for four
arbitrary surface functions which were chosen to be the  plasma
density, axial Mach function, poloidal Mach-function and static
pressure.

 After assigning  the free  functions,  a systematic examination of
equilibrium quantities and confinement figures of merit, as the
safety factor,  electric field  and ${\bf E}\times {\bf B} $
velocity shear for a variety of parametric values,  implies  that
the equilibrium characteristics are qualitatively consistent with
experimental evidence of the L-H transition. In addition,
application of a sufficient condition for linear stability and
parallel  flow indicates that  one statioanry  equilibrium being
potential stable may describe the H-state and another static equilibrium
not satisfying  the stability condition the L-state. In addition the
equilibrium non-linearity in conjunction with the flow and the flow
shear may play a stabilizing role. Although understanding the
physics of the L-H transition remains incomplete  the results of the
present study may  shed some light towards  that goal.

Finally it would be interesting trying to generalize these classes
of solutions to up-down asymmetric configuration with a lower $x$-
point in connection with the ITER project. Also the study could be
extended to toroidal geometry in order to examine the impact of
toroidicity.


\section*{Aknowledgments}\

One of the authors (GNT) would like to thank Drs. Henri Tasso and
Calin Atanasiu for very useful discussions.

The work leading to this article was performed within
the participation of the University of Ioannina in the Association
Euratom-Hellenic Republic, which is supported in
part by the European Union (Contract of Association No.
ERB 5005 CT 99 0100) and by the General Secretariat of
Research and Technology of Greece. The views and opinions
expressed herein do not necessarily reflect those of the European
Commission.


\end{document}